\documentclass[oldversion]{aa}
\usepackage{graphicx}
\usepackage{graphics}
\usepackage{epsfig}

\newcommand{\be}{\begin{eqnarray}}
\newcommand{\ee}{\end{eqnarray}}

\begin{document}

\title{A simplified model of jet power from active galactic nuclei}
\author{Yang Li, Ding-Xiong Wang and Zhao-Ming Gan}
\titlerunning{A simplified model of jet power from active galactic
nuclei}
\authorrunning{Yang Li, Ding-Xiong Wang and Zhao-Ming Gan}

\institute{Department of Physics, Huazhong University of Science
and Technology, Wuhan, 430074, China}

\offprints{D. X. Wang\\  \email{dxwang@mail.hust.edu.cn}}
\date{Received  / Accepted }

\abstract{
 \textbf{Aims. }A simplified model of jet power from
active galactic nuclei is proposed in which the relationship between
jet power and disk luminosity is discussed by combining disk
accretion with two mechanisms of extracting energy magnetically from
a black hole accretion disk, i.e., the Blandford-Payne (BP) and the
Blandford-Znajek (BZ) processes.

\textbf{Methods.} By including the BP process into the conservation
laws of mass, angular momentum and energy, we derive the expressions
of the BP power and disk luminosity, and the jet power is regarded
as the sum of the BZ and BP powers.

\textbf{Results. }We find that the disk radiation flux and
luminosity decrease because a fraction of the accretion energy is
channelled into the outflow/jet in the BP process. It is found that
the dominant cooling mode of the accretion disk is determined mainly
by how the poloidal magnetic field decreases with the cylindrical
radius of the jet. By using the parameter space we found, which
consists of the black hole spin and the self-similar index of the
configuration of the poloidal magnetic field frozen in the disk, we
were able to compare the relative importance of the following
quantities related to the jet production: (\ref{eq1}) the BP power
versus the disk luminosity, (\ref{eq2}) the BP power versus the BZ
power, and (\ref{eq3}) the jet power versus the disk luminosity. In
addition, we fit the jet power and broad-line region luminosity of
11 flat-spectrum radio quasars (FSRQs) and 17 steep-spectrum radio
quasars (SSRQs) based on our model.}


\maketitle

\section{Introduction}

Much attention has been paid to the relativistic jet and the
enormous amounts of energy released in active galactic nuclei (AGNs)
in the past decades. It is widely believed that the Blandford-Znajek
(BZ) process (Blandford {\&} Znajek 1977, hereafter BZ77; Macdonald
{\&} Thorne 1982) and the Blandford-Payne (BP) process (Blandford
{\&} Payne 1982, hereafter BP82; Spruit 1996, hereafter S96) are the
major mechanisms powering the relativistic jet from AGN hosting a
supermassive black hole.

Energy and angular momentum are extracted from a rotating black hole
to power the jet in the BZ process, in which the poloidal magnetic
field lines connecting the black hole horizon with remote
astrophysical loads are invoked. In the BP process, the disk matter
is channelled into the outflow/jet by virtue of the poloidal
magnetic field lines frozen in the disk, and the streaming gas is
accelerated due to the work done by the magnetic torque. It has been
argued that the kinetic flux carried by the outflow/jet driven
centrifugally in the BP process always accompanies the Poynting flux
(BP82; Camenzind 1986; S96).

Maraschi {\&} Tavecchio (2003, hereafter MT03) discuss the relation
between the power carried by relativistic jets and the nuclear power
provided by accretion for a group of blazars, including
flat-spectrum radio quasars (FSRQs) and BL Lac objects. Their
analysis indicates that the total jet power is of the same order of
magnitude as the accretion power for FSRQs, while the jet luminosity
is higher than the disk luminosity for BL Lac objects. The same
result has been obtained by D'Elia, Padovani {\&} Landt (2003).

Very recently, Liu, Jiang {\&} Gu (2006, hereafter L06) investigated
the relation between the jet power and the black hole mass in
radio-loud AGNs. In their work, the jet power was estimated by using
extrapolated, extended 151 MHz flux density from the VLA 5 GHz
extended radio emission based on the formula derived by Punsly
(2005), and the broad-line region luminosity and the black hole mass
can be estimated by the broad emission-line luminosity (Celotti et
al.1997; McLure {\&} Dunlop 2001; McLure {\&} Jarvis 2002).

On the other hand, Miller et al. (2006) stress that disk accretion
onto black holes is a fundamentally magnetic process: internal
viscosity in some magnetic processes and disk winds can transfer
angular momentum to drive disk accretion. It has been pointed out
that an outflow emanating from an accretion disk can act as a sink
for mass, angular momentum, and energy, altering the dissipation
rates and effective temperatures across the disk (Donea {\&}
Biermann 1996; Knigge 1999; Kuncic {\&} Bicknell 2007).

Motivated by the above works, we discuss the outflow/jet driven by
the BP process, and investigate the interaction of the outflow/jet
with the disk accretion based on the conservation laws of mass,
angular momentum and energy. We find that the disk radiation flux
and luminosity are reduced due to a fraction of accretion energy
being channelled into the outflow/jet by the poloidal magnetic field
frozen in the disk. It is shown that the dominant cooling mode in
the disk is determined by the black hole spin $a_\ast $ and by the
self-similar index $\alpha $ for the fixed-jet Lorentz factor. In
addition, we find that the BP power is generally greater than the BZ
power, except when the black hole spins very fast and the magnetic
field decreases very steeply with the cylindrical radius. We compare
the jet power and the disk luminosity and find that the jet power is
almost the same as the disk luminosity. In this model the jet power
is regarded as the sum of the BZ and BP powers, and the broad-line
region luminosity is taken as a fraction of disk luminosity. In this
way, 11 FSRQs and 17 steep-spectrum radio quasars (SSRQs) are
fitted, and these results are consistent with those given in L06.

This paper is organized as follows. In Sect. 2 we describe our model
and discuss the accretion rate and the radiation flux at the
presence of a jet based on the conservation laws of mass, angular
momentum, and energy. In Sect. 3 we compare the importance of the BP
power to the disk luminosity, the BP power to the BZ power and the
jet power to the disk luminosity in the parameter space consisting
of the black hole spin and the self-similar index of the
configuration of the poloidal magnetic field frozen in the disk. In
addition, we fit the jet power and broad-line region luminosity of
11 FSRQs and 17 SSRQs. Finally, in Sect. 4, we summarize our main
results and discuss the limitation of our model. Throughout this
paper the units $G=c=$1  are used.

\section{DESCRIPTION OF OUR MODEL}

As is well known, large-scale magnetic fields anchored in the black
hole accretion disk play essential roles in jet formation (Blandford
2002). Two scenarios have been proposed to interpret the origin of
the large-scale magnetic fields. One is based on the results of some
numerical simulations, indicating that small-scale magnetic fields
could be amplified by virtue of a dynamo process in accretion disks
(Hawley et al. 1995; Tout {\&} Pringle 1996; Amitage 1998). However,
S96 thought that this would not be the ideal field for driving
magnetic winds. Another possibility is that the magnetic field could
be captured and advected inwards by the accreting matter in disks
(BP82; Lovelace 1994; Spruit et al. 2005). The trapped large-scale
fields can be strong enough to produce magnetic outflows.

In this paper we assume that the accretion disk is thin, Keplerian,
stable, and perfectly conducting, located in an equatorial plane of
a rotating black hole, and the inner edge of the disk is the last
stable circular orbit (ISCO, Novikov {\&} Thorne 1973). The magnetic
field configuration is assumed to be as shown in Fig. 1.

\begin{figure}
\centering
\centerline{\includegraphics[width=7cm]{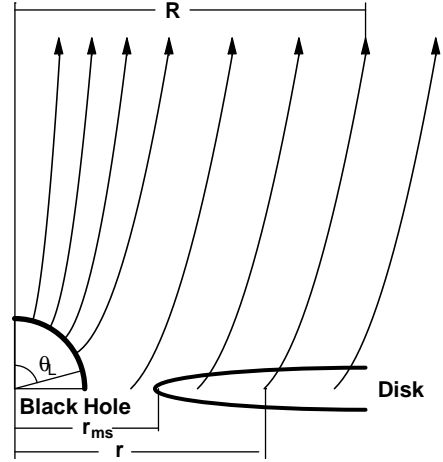}}
\caption[]{Configuration of poloidal magnetic field threading a
rotating black hole and its surrounding disk. We take
$\theta_{L}=0.5\pi$ throughout this paper.} \label{fig1}
\end{figure}

Following BP82, we assume that the poloidal magnetic field on the
disk surface varies with the disk radius as

\begin{equation}
\label{eq1} B_D^p = B_H^p(r/r_{H})^{-5/4},
\end{equation}

\noindent where $r$ is the disk radius and $r_H = M(1 + \sqrt {1 -
a_\ast ^2 } )$ is the horizon radius of the black hole. The
quantities $B_D^p $ and $B_H^p $ are the poloidal magnetic field
at the disk surface and black hole horizon, respectively.

The poloidal magnetic field far from the disk surface is assumed to
be roughly self-similar, being given as (BP82, Lubow et al. 1994)

\begin{equation}
\label{eq2} B^p \sim B_D^p \left( {R \mathord{\left/ {\vphantom {R
r}} \right. \kern-\nulldelimiterspace} r} \right)^{ - \alpha },
\quad \left( {\alpha \ge 1} \right)
\end{equation}

\noindent where $\alpha $ is the self-similar index to describe
the variation of the poloidal magnetic field with the cylindrical
radius $R$ of the jet.

Considering the balance between the magnetic pressure on the horizon
and the ram pressure in the innermost parts of an accretion flow,
Moderski, Sikora {\&} Lasota (1997) expressed the magnetic field at
the horizon as

\begin{equation}
\label{eq3} {\left( {B_H^p } \right)^2} \mathord{\left/ {\vphantom
{{\left( {B_H^p } \right)^2} {8\pi }}} \right.
\kern-\nulldelimiterspace} {8\pi } = P_{ram} \sim \rho c^2\sim
{\dot {M}_{acc} (r_{ms} )} \mathord{\left/ {\vphantom {{\dot
{M}_{acc} (r_{ms} )} {\left( {4\pi r_H^2 } \right)}}} \right.
\kern-\nulldelimiterspace} {\left( {4\pi r_H^2 } \right)},
\end{equation}

\noindent where $\dot {M}_{acc} (r_{ms} )$ is the accretion rate
at ISCO with the radius $r_{ms} $. Eq. (\ref{eq3}) can be
rewritten as

\begin{equation}
\label{eq4} \dot {M}_{acc} (r_{ms} ) \approx {\left( {B_H^p }
\right)^2r_H^2 } \mathord{\left/ {\vphantom {{\left( {B_H^p }
\right)^2r_H^2 } 2}} \right. \kern-\nulldelimiterspace} 2.
\end{equation}

The electromagnetic outflow/jet can be driven by the BZ process, in
which the energy is extracted from a spinning black hole. The
optimal BZ power is given (Wang et al. 2002)

\begin{equation}
\label{eq5} {P_{BZ} } \mathord{\left/ {\vphantom {{P_{BZ} } {P_0
}}} \right. \kern-\nulldelimiterspace} {P_0 } = 2{A^{ - 1}\left(
{\arctan A - {a_\ast } \mathord{\left/ {\vphantom {{a_\ast } 2}}
\right. \kern-\nulldelimiterspace} 2} \right)} \mathord{\left/
{\vphantom {{A^{ - 1}\left( {\arctan A - {a_\ast } \mathord{\left/
{\vphantom {{a_\ast } 2}} \right. \kern-\nulldelimiterspace} 2}
\right)} {(1 + q)^2}}} \right. \kern-\nulldelimiterspace} {(1 +
q)^2},
\end{equation}

\noindent where $a_\ast $ is the black hole spin, and the quantities
$P_0 $, $A$, and $q$ are defined as $P_0 \equiv \dot {M}_{acc}
(r_{ms} )c^2$, $A \equiv \sqrt {(1 - q) / (1 + q)} $, and $q \equiv
\sqrt {1 - a_\ast ^2 } $, respectively.

As argued in BP82, the outflow matter could be accelerated
centrifugally along the magnetic field lines, overcoming a barrier
of gravitational potential to form magnetohydrodynamic (MHD) jets,
provided that the poloidal magnetic field is strong and inclined
enough. To avoid complexity in jet acceleration, Cao (2002,
hereafter C02) expresses the mass loss rate in the jet from unit
area of the disk surface as

\begin{equation}
\label{eq6} \dot {m}_{jet} = \frac{\left( {B_D^p } \right)^2}{4\pi
}\frac{\left[ {r\Omega _D } \right]^\alpha \gamma _j^\alpha
}{\left( {\gamma _j^2 - 1} \right)^{{(1 + \alpha )}
\mathord{\left/ {\vphantom {{(1 + \alpha )} 2}} \right.
\kern-\nulldelimiterspace} 2}},
\end{equation}

\noindent where $\gamma _j $ is the Lorentz factor of the jet. The
quantity $\Omega _D $ is the Keplerian angular velocity at the foot
point of the field line:

\begin{equation}
\label{eq7} \Omega _D = \frac{1}{M\left( {\xi ^{3 / 2}\chi _{ms}^3
+ a_ * } \right)},
\end{equation}

\noindent where $\xi \equiv r \mathord{\left/ {\vphantom {r {r_{ms}
}}} \right. \kern-\nulldelimiterspace} {r_{ms} }$ is a radial
parameter of the disk defined in terms of the radius $r_{ms} $, and
$\chi _{ms} $ is defined as $\chi _{ms} \equiv \sqrt {{r_{ms} }
\mathord{\left/ {\vphantom {{r_{ms} } M}} \right.
\kern-\nulldelimiterspace} M} $.

According to the mass conservation law, the accretion rate of disk
matter is related to the mass outflow rate by

\begin{equation}
 \label{eq8} {d\dot {M}_{acc} (r)} \mathord{\left/ {\vphantom
{{d\dot {M}_{acc} (r)} {dr}}} \right. \kern-\nulldelimiterspace}
{dr} = 4\pi r\dot {m}_{jet} (r),
\end{equation}

\noindent where $\dot {M}_{acc} (r)$ and $\dot {m}_{jet} (r)$ are
the accretion rate and the mass loss rate at foot point,
respectively. Integrating Eq. (\ref{eq8}), we have

\begin{equation}
\label{eq9} \dot {M}_{acc} (r) = \dot {M}_{acc} (r_{ms} ) +
\int_{r_{ms} }^r {4\pi {r}'\dot {m}_{jet} d{r}'} ,
\end{equation}

\noindent where $\dot {M}_{acc} (r_{ms} )$ is the accretion rate
at ISCO.

Incorporating Eqs. (\ref{eq1}), (\ref{eq4}), (\ref{eq6}),
(\ref{eq7}), and (\ref{eq9}), we have the accretion rate

\be \label{eq10} \dot {m}_{acc} (a_\ast ,\alpha ,\gamma _j ,\xi )
&=& \dot {M}_{acc} (r)/\dot {M}_{acc} (r_{ms} ) \nonumber\\&=& 1 +
2\int_1^\xi {g_{jet} {\xi }'^{\alpha - 3 \mathord{\left/
{\vphantom {3 2}} \right. \kern-\nulldelimiterspace} 2}d{\xi }'} ,
\ee

\noindent where $g_{jet} $ is defined as\\
$g_{jet} = {\left( {r_{ms} \Omega _D } \right)^\alpha \xi _H^{1
\mathord{\left/ {\vphantom {1 2}} \right.
\kern-\nulldelimiterspace} 2} \gamma _j^\alpha } \mathord{\left/
{\vphantom {{\left( {r_{ms} \Omega _D } \right)^\alpha \xi _H^{1
\mathord{\left/ {\vphantom {1 2}} \right.
\kern-\nulldelimiterspace} 2} \gamma _j^\alpha } {\left( {\gamma
_j^2 - 1} \right)^{{(1 + \alpha )} \mathord{\left/ {\vphantom {{(1
+ \alpha )} 2}} \right. \kern-\nulldelimiterspace} 2}}}} \right.
\kern-\nulldelimiterspace} {\left( {\gamma _j^2 - 1} \right)^{{(1
+ \alpha )} \mathord{\left/ {\vphantom {{(1 + \alpha )} 2}}
\right. \kern-\nulldelimiterspace} 2}}$.

From Eq. (\ref{eq10}) we find that the accretion rate at the given
radius is determined by three parameters: the self-similar index
$\alpha $, the Lorentz factor $\gamma _j $, and the black hole spin
$a_\ast $. Based on Eq. (\ref{eq10}) we have the curves of the
accretion rate $\dot {m}_{acc} $ varying with the radial parameter
$\xi $ for the given values of $\alpha $, $a_\ast $, and $\gamma _j
$ as shown in Fig. 2. It is shown that $\dot {m}_{acc} $ increases
very steeply with the increasing $\xi $ in the innermost region of
the disk, while it almost stays constant because the disk radius is
greater than several $r_{ms} $. This result implies that the outflow
is launched predominantly from the innermost region of the disk.

\begin{figure}
\centering
\centerline{\includegraphics[width=7cm]{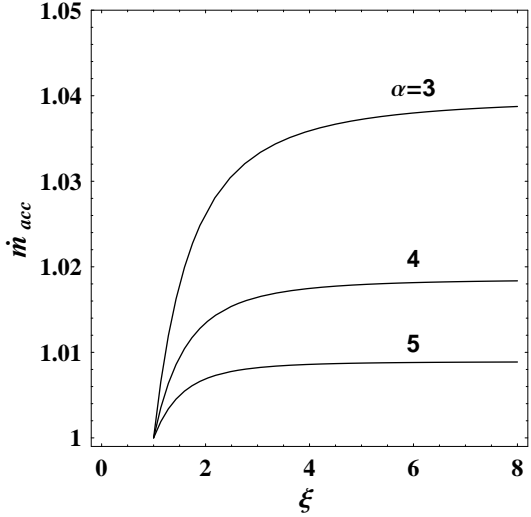}}
\caption[]{The curves of $\dot {m}_{acc} $ versus $\xi $ for
different values of $\alpha $ with $a_\ast = 0.95$ and $\gamma _j =
5$.} \label{fig2}
\end{figure}

The ratio of the total mass-loss rate in the outflow to the
accretion rate at ISCO is defined as

\begin{equation} \label{eq11} \eta = \frac{\dot {M}_{acc} (r_{out} ) - \dot
{M}_{acc} (r_{ms} )}{\dot {M}_{acc} (r_{ms} )} = 2\int_1^{\xi
_{out} } {g_{jet} {\xi }'^{\alpha - 3 \mathord{\left/ {\vphantom
{3 2}} \right. \kern-\nulldelimiterspace} 2}d{\xi }'} ,
\end{equation}

\noindent where $\xi _{out} \equiv {r_{out} } \mathord{\left/
{\vphantom {{r_{out} } {r_{ms} }}} \right.
\kern-\nulldelimiterspace} {r_{ms} }_{ }$is the dimensionless outer
radius $r_{out} $ of the jet. For different values of $\alpha $, the
curves of $\lg \eta $ versus $\gamma _j $ with the given $a_ * $,
and those of $\lg \eta $ versus $a_ * $ with the given $\gamma _j $
are shown in Figs. 3a and 3b, respectively. It is shown in Fig. 3
that the ratio of the total mass-loss rate to the accretion rate
decreases monotonically with increasing $\alpha $ and $\gamma _j $,
while it increases monotonically with increasing $a_\ast $. These
results are consistent with those of C02.

\begin{figure}[htbp]
\centerline{\includegraphics[width=70mm,height=70mm]{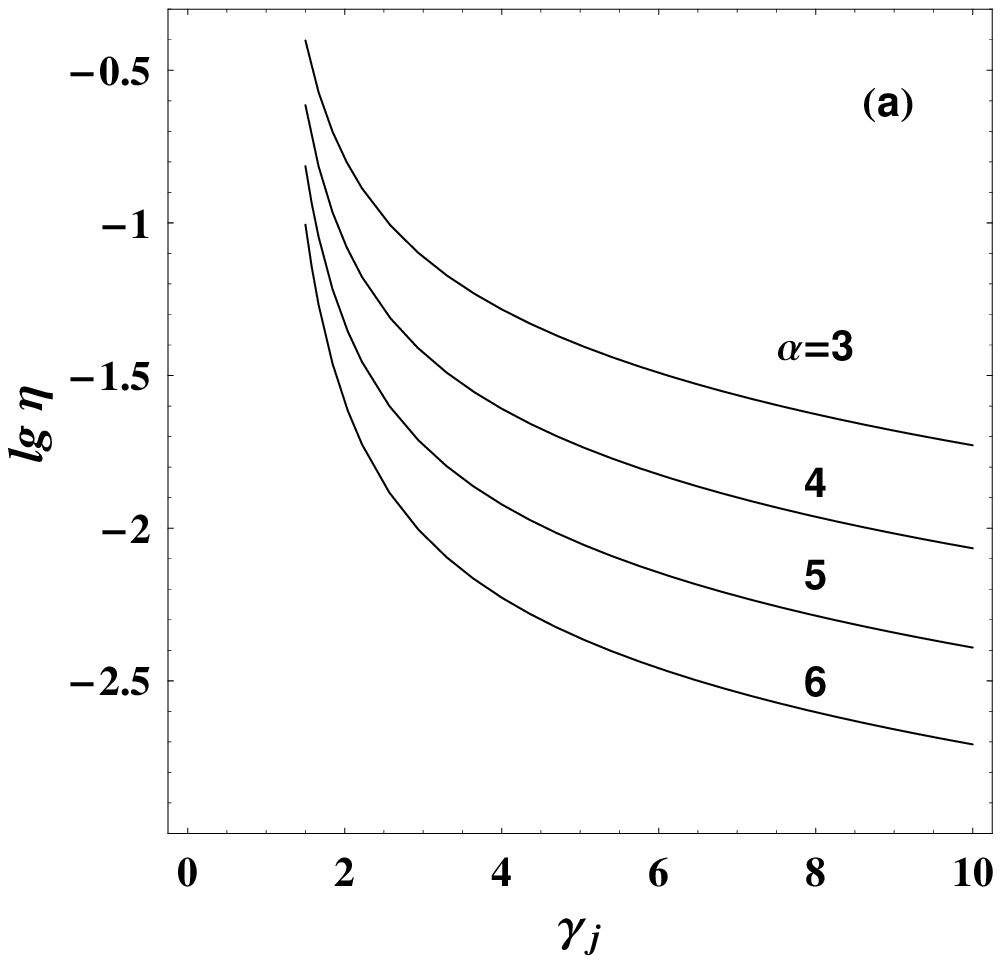}}
\centerline{\includegraphics[width=70mm,height=70mm]{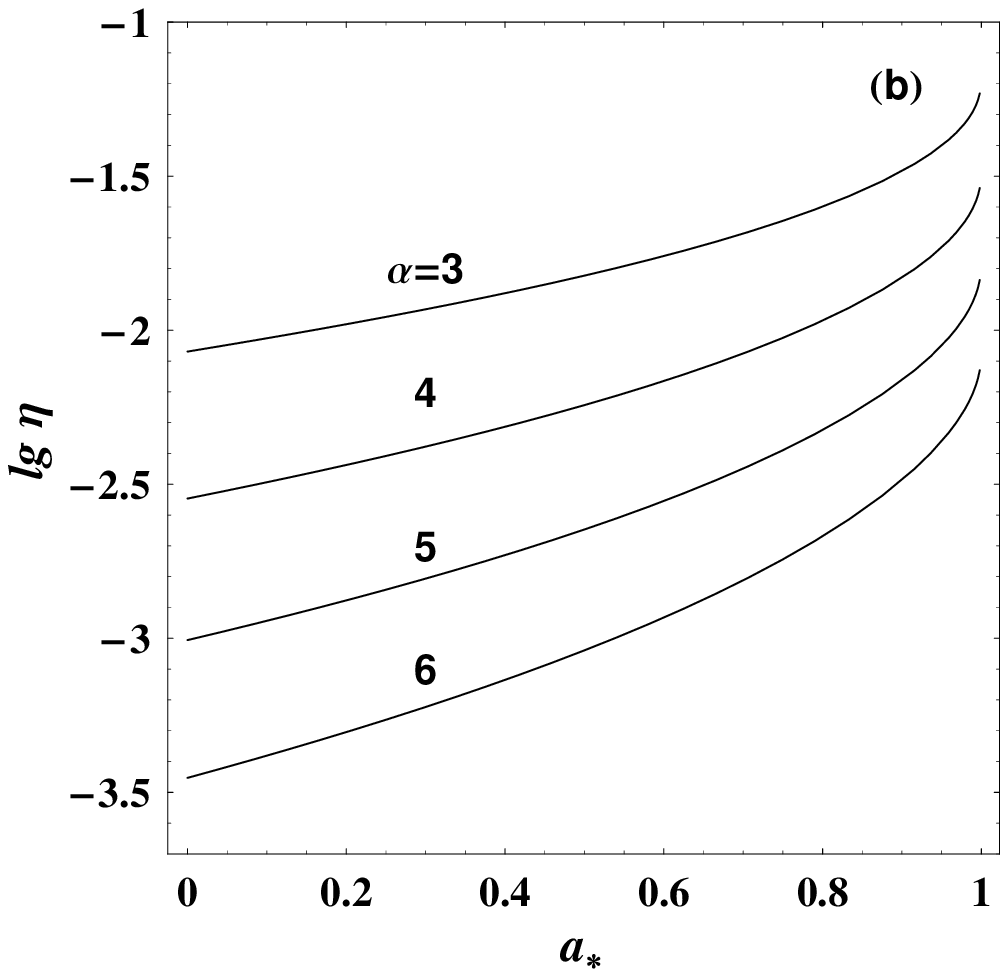}}
\caption[]{The curves of $\lg \eta $ versus $\gamma _j $ for $a_\ast
= 0.95$, (b) the curves of $\lg \eta $ versus $a_\ast $ for $\gamma
_j = 5$. In both cases $\alpha $= 3, 4, 5, 6 and $\xi _{out} = 10^4$
are assumed.} \label{fig3}
\end{figure}

Camenzind (1986) pointed out that the Poynting flux, as well as the
kinetic flux, are carried by the winds driven centrifugally from the
supermassive objects, and a fraction of the electromagnetic energy
and angular momentum extracted is converted into the kinetic energy
of matter in the outflow. It has been argued in BP82 and S96 that
the Poynting flux dominates the kinetic flux near the disk surface,
while the former is converted into the latter during accelerating
matter in the outflow. Based on the calculations in BP82, the ratio
of the Poynting flux to the kinetic flux is about 58 near the disk
surface, while it reduces to 2 at the Alfven surface. This result
implies that about one third of the energy in the Poynting flux has
been converted into the kinetic energy in the driving process.

Following C02, the kinetic flux of the jet can be written as

\begin{equation}
\label{eq12} F_{jet} = \dot {m}_{jet} c^2(\gamma _j - 1).
\end{equation}

\noindent Considering that the Poynting flux is much greater than
the kinetic flux near the disk surface, and about one third of the
energy in the Poynting flux is converted into the kinetic energy of
the jet, we can relate $F_{jet} $ at Alfven surface to the Poynting
flux at the disk surface:

\begin{equation}
\label{eq13} S_E = 3F_{jet} .
\end{equation}

According to BP82, the jet is driven from the disk surface due to
the work done by the magnetic torques, and the angular momentum flux
$S_L $ extracted electromagnetically from the disk surface is
related to the Poynting energy flux as

\begin{equation}
\label{eq14} S_L \mbox{ = }{S_E } \mathord{\left/ {\vphantom {{S_E
} {\Omega _D }}} \right. \kern-\nulldelimiterspace} {\Omega _D }.
\end{equation}

\noindent Incorporating Eqs. (\ref{eq12})---(\ref{eq14}), we have

\begin{equation}
\label{eq15} S_L \mbox{ = }{3\dot {m}_{jet} (\gamma _j - 1)}
\mathord{\left/ {\vphantom {{3\dot {m}_{jet} (\gamma _j - 1)}
{\Omega _D }}} \right. \kern-\nulldelimiterspace} {\Omega _D }.
\end{equation}

The integrated shear stress of the disk should be unavoidably
affected by the transport of angular momentum and energy in the jet,
resulting in the decrease of the disk dissipation and disk
radiation. Unfortunately, this fact is neglected in C02, in which
the strength of the large-scale field threading the disk is
estimated based on dynamo mechanisms in the accretion disk and the
expression of the integrated shear stress given by Novikov {\&}
Thorne (1973).

At the appearance of the jet, the conservation equations of energy
and angular momentum can be written as

 \[
\frac{d}{dr}\left( {\dot {M}_{acc} E^\dag - T_{visc} \Omega _D }
\right)
\]

\begin{equation}
\label{eq16} \ \ \ \ \ \ \ \  = 4\pi r\left[ {\left( {\dot
{m}_{jet} + F_{rad} } \right)E^\dag + S_L \Omega _D } \right],
\end{equation}

\begin{equation}
\label{eq17} \frac{d}{dr}\left( {\dot {M}_{acc} L^\dag - T_{visc}
} \right) = 4\pi r\left[ {\left( {\dot {m}_{jet} + F_{rad} }
\right)L^\dag + S_L } \right],
\end{equation}

\noindent where $T_{visc} $ and $F_{rad} $ are the internal
viscous torque and the energy flux radiated away from the surface
of disk, respectively.

In Eqs. (\ref{eq16}) and (\ref{eq17}), $E^\dag $ and $L^\dag $ are
the specific energy and angular momentum of the disk matter, being
expressed by (Novikov {\&} Thorne 1973)

\begin{equation}
\label{eq18} E^\dag = {\left( {1 - 2\chi ^{ - 2} + a_ * \chi ^{ -
3}} \right)} \mathord{\left/ {\vphantom {{\left( {1 - 2\chi ^{ -
2} + a_ * \chi ^{ - 3}} \right)} {\left( {1 - 3\chi ^{ - 2} + 2a_
* \chi ^{ - 3}} \right)^{1 / 2}}}} \right.
\kern-\nulldelimiterspace} {\left( {1 - 3\chi ^{ - 2} + 2a_ * \chi
^{ - 3}} \right)^{1 / 2}},
\end{equation}

\be \label{eq19} &L^\dag =& \nonumber\\&M\chi&{\left( {1 - 2a_ *
\chi ^{ - 3} + a_ * ^2 \chi ^{ - 4}} \right)} \mathord{\left/
{\vphantom {{\left( {1 - 2a_ * \chi ^{ - 3} + a_ * ^2 \chi ^{ -
4}} \right)} {\left( {1 - 3\chi ^{ - 2} + 2a_ * \chi ^{ - 3}}
\right)^{1 / 2}}}} \right. \kern-\nulldelimiterspace} {\left( {1 -
3\chi ^{ - 2} + 2a_ * \chi ^{ - 3}} \right)^{1 / 2}}. \ee

\noindent where $\chi \equiv \sqrt {r \mathord{\left/ {\vphantom
{r M}} \right. \kern-\nulldelimiterspace} M} = \xi ^{1
\mathord{\left/ {\vphantom {1 2}} \right.
\kern-\nulldelimiterspace} 2}\chi _{ms} $, and the quantities
$L^\dag $ and $E^\dag $ are related by

\begin{equation}
\label{eq20} \frac{dE^\dag }{dr} = \Omega _D \frac{dL^\dag }{dr}.
\end{equation}

The terms on the left hand side of Eqs. (\ref{eq16}) and
(\ref{eq17}) are the radial transfer of energy and angular momentum
due to disk accretion, respectively, while the terms on the
right-hand side of these equations represent the transfer of energy
and angular momentum due to disk radiation and jet.

Not long ago, some authors (Balbus {\&} Hawley 1998; Agol {\&}
Krolik 2000) pointed out that the magnetic stresses might exert a
time-steady torque on the inner edge of the disk, and a nonzero
torque at $r_{ms} $ can be expressed as follows,

\begin{equation}
\label{eq21} T_{ms} \approx \frac{\left( {B_{ms}^p }
\right)^2}{4\pi }4\pi H_{ms} r_{ms}^2 = 0.2\dot {M}_{acc} (r_{ms}
)\sqrt {r_{ms} r_H } ,
\end{equation}

\noindent where $B_{ms}^p $, $T_{ms} $, and $H_{ms} $ are the
poloidal magnetic field, torque, and the height of the disk at the
inner edge of disk, respectively, and $\left( {H \mathord{\left/
{\vphantom {H {r_{ms} }}} \right. \kern-\nulldelimiterspace} {r_{ms}
}} \right)_{\max } \approx 0.1$ is assumed in calculations.
Equations (\ref{eq1}) and (\ref{eq4}) are used in the last step of
deriving Eq. (\ref{eq21}).

Incorporating Eqs. (\ref{eq16})---(\ref{eq20}), we have

\[
F_{rad} = - \frac{{d\Omega _D } \mathord{\left/ {\vphantom
{{d\Omega _D } {dr}}} \right. \kern-\nulldelimiterspace}
{dr}}{4\pi r}(E^\dag - \Omega _D L^\dag ))^{ - 2}
\]

\[
\times \left[ {\left( {\int_{r_{ms} }^r {(E^\dag - \Omega _D
L^\dag )\left( {\dot {M}_{acc} \frac{dL^ + }{dr}} \right) +
(E_{ms}^\dag - \Omega _{ms} L_{ms}^\dag )T_{ms} } } \right)}
\right.
\]

\begin{equation}
\label{eq22} \left. { - \int_{r_{ms} }^r {(E^\dag - \Omega _D
L^\dag )4\pi rS_L } dr} \right].
\end{equation}

\noindent The quantity $F_{rad} $ in Eq. (\ref{eq22}) is the disk
radiation flux in the presence of the jet, while the first integral
on the right hand side of the equation represents the release rate
of the accreting matter's energy, in which a magnetic torque exerted
at ISCO is taken into account, and the second integral is the
cooling rate due to the outflow/jet driven by the BP process, in
which the kinetic flux and Poynting flux are included. It is
expected that some relation between the BP power and disk luminosity
can be obtained based on our model.

\section{ RELATIONSHIP BETWEEN JET POWER AND DISK LUMINOSITY }

The relationship between the BP power and disk luminosity can be
discussed based on Eq. (\ref{eq22}), and the curves of ${F_{rad} }
\mathord{\left/ {\vphantom {{F_{rad} } {F_0 }}} \right.
\kern-\nulldelimiterspace} {F_0 }$ versus $\xi $ for the given
values of $a_ * $, $\gamma _j $, and $\alpha $ are shown in Fig. 4,
in which the disk radiation flux is significantly reduced due to the
existence of the jet driven by the BP process. This result is
consistent with ones from other authors (Donea {\&} Biermann 1996;
Knigge 1999; Kuncic {\&} Bicknell 2007).

\begin{figure}[htbp]
\centerline{\includegraphics[width=70mm,height=70mm]{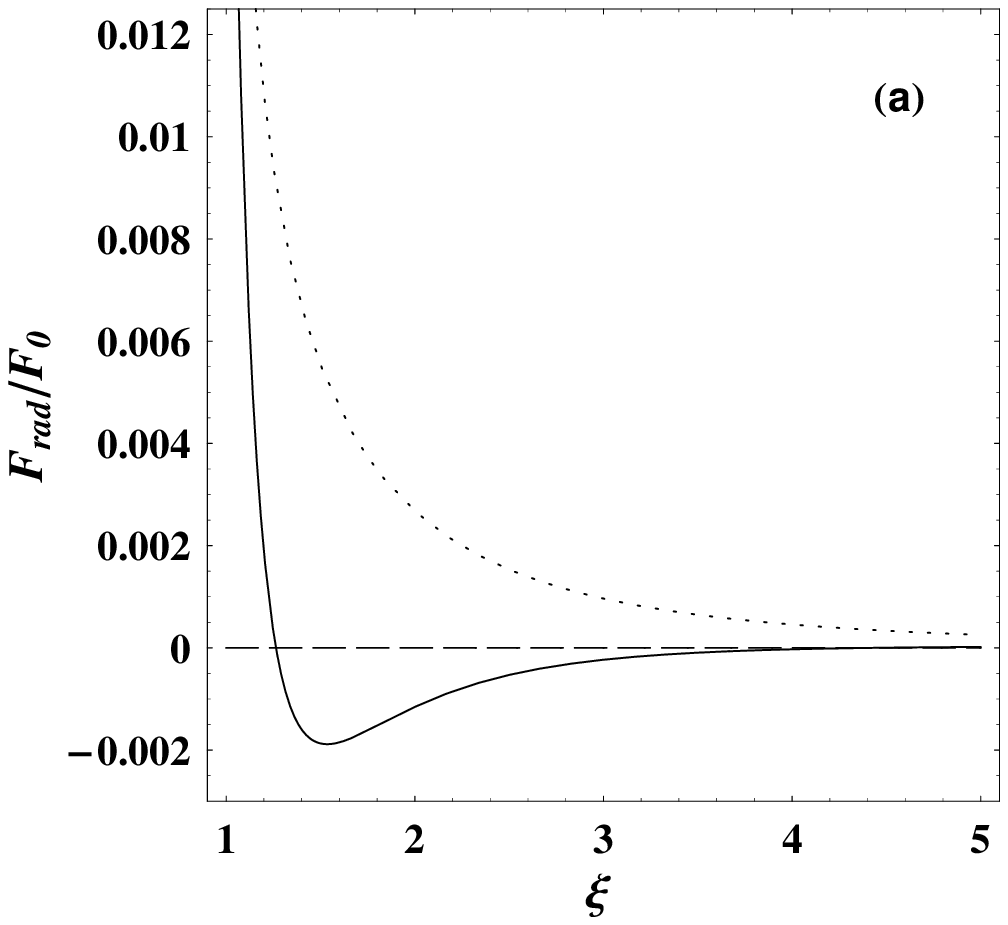}}
\centerline{\includegraphics[width=70mm,height=70mm]{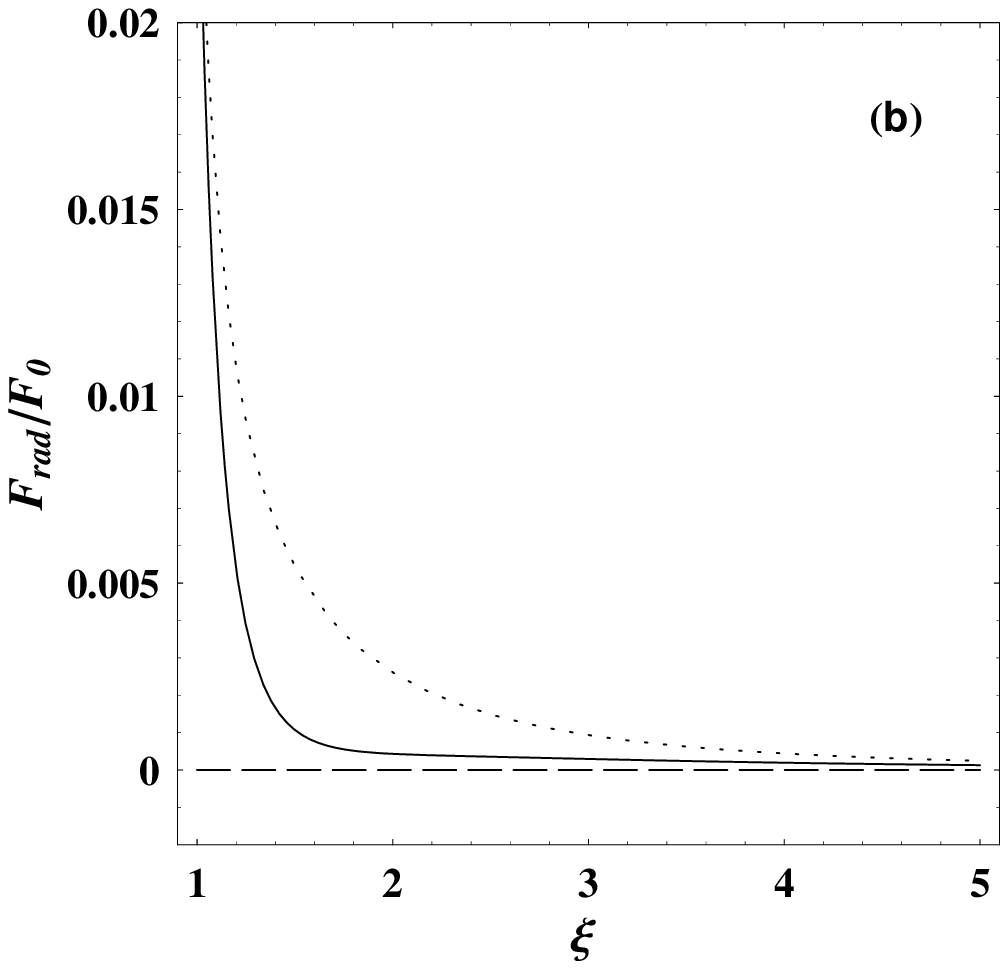}}
\caption[]{The curves of $F_{rad}/F_{0}$  versus  $\xi $ for (a)
$a_{*} = 0$ and $\alpha = 4$; (b) $a_{*} = 0.95$  and $\alpha = 5$ .
The solid and dotted lines represent the disk radiation flux with
and without jet, respectively. The parameter
$F_{0}=\dot{M}_{acc}(r_{ms})/(r_{ms}^{2}M^{2})$ is defined, and
$\gamma_{j}=10$ are assumed.} \label{fig4}
\end{figure}

From Fig. 4a we find that the radiation flux at the presence of the
jet could become negative in the inner disk, and this result is
unphysical. Inspecting Eq. (\ref{eq21}), we find that this
unphysical result can be removed, provided that the cooling
contribution of the BP process is not very strong. The following
condition is required by non-negative radiation flux:

\begin{equation}
\label{eq23} \left( {F_{rad} } \right)_{\min } \ge \mbox{0},
\end{equation}

\noindent where $\left( {F_{rad} } \right)_{\min } $ is the
minimum disk radiation at the presence of the jet.

Combining Eq. (\ref{eq23}) with Eq. (\ref{eq22}), we have the
contour surface of $\left( {F_{rad} } \right)_{\min } = \mbox{0}$
in the 3-dimensional parameter space $\left( {a_ * ,\alpha ,\gamma
_j } \right)$ as shown in Fig. 5, in which the values of the
parameters above the contour surface $\left( {F_{rad} }
\right)_{\min } = \mbox{0}$ correspond to positive radiation flux,
while those below the surface are unphysical.

\begin{figure}
\centering
\centerline{\includegraphics[width=70mm,height=70mm]{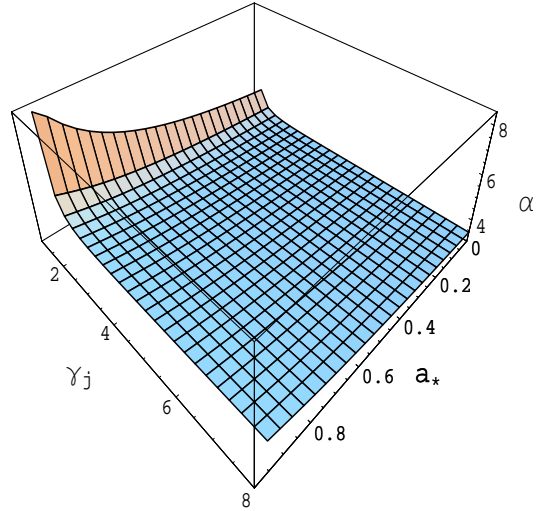}}
\caption[]{The contour surface of $\left( {F_{rad} } \right)_{\min }
= \mbox{0}$ in the 3-dimensional parameter space $\left( {a_ *
,\alpha ,\gamma _j } \right)$.} \label{fig5}
\end{figure}

Inspecting Fig. 5, we find that the requirement for non-negative
radiation flux is very sensitive to the self-similar index $\alpha
$, which should be greater than some critical value for the given
black hole spin $a_ * $ and Lorentz factor $\gamma _j $ of the
jet. This result implies that the poloidal magnetic field
expressed by Eq. (\ref{eq2}) should reduce steeply enough with the
increasing cylindrical radius to avoid an unphysical disk flux.

Taking the nonzero torque exerted at $r_{ms} $ into account, we have
the disk luminosity by integrating Eq. (\ref{eq16}) as

\be \label{eq24} L_{disk} &=& \int_{r_{ms} }^{r_{out} } {4\pi
rF_{rad} E^\dag dr} = \int_{r_{ms} }^{r_{out} } {\dot {M}_{acc}
(r)dE^\dag } \nonumber \\&-& 4\pi \int_{r_{ms} }^{r_{out } } {S_L
\Omega _D rd} r + T_{ms} \Omega _D . \ee

\noindent And the BP power can be expressed as

\begin{equation}
\label{eq25} P_{BP} = \int_{r_{ms} }^{r_{out} } {4\pi r\left(
{\dot {m}_{jet} E^\dag + S_L \Omega _D } \right)dr} ,
\end{equation}

\noindent where the kinetic and electromagnetic energy are included.
Inspecting Eqs. (\ref{eq24}) and (\ref{eq25}), we find that the
ratio of the BP power to the disk luminosity depends on the
parameters $a_ * $, $\alpha $, and $\gamma_{j} $, and the curves of
$\log \left( {{P_{BP} } \mathord{\left/ {\vphantom {{P_{BP} }
{L_{\mbox{disk}} }}} \right. \kern-\nulldelimiterspace}
{L_{\mbox{disk}} }} \right)$ versus $\gamma_{j} $ with different
values of $a_ * $ and $\alpha $ are shown in Fig. 6.

\begin{figure}[htbp]
\centerline{\includegraphics[width=70mm,height=70mm]{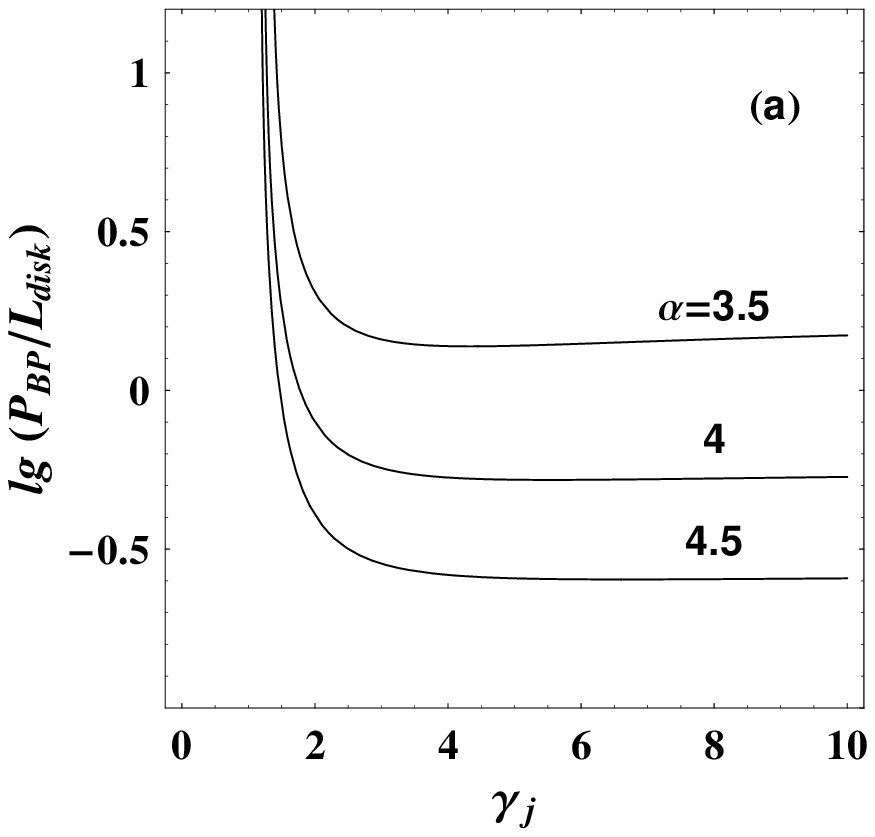}}
\centerline{\includegraphics[width=70mm,height=70mm]{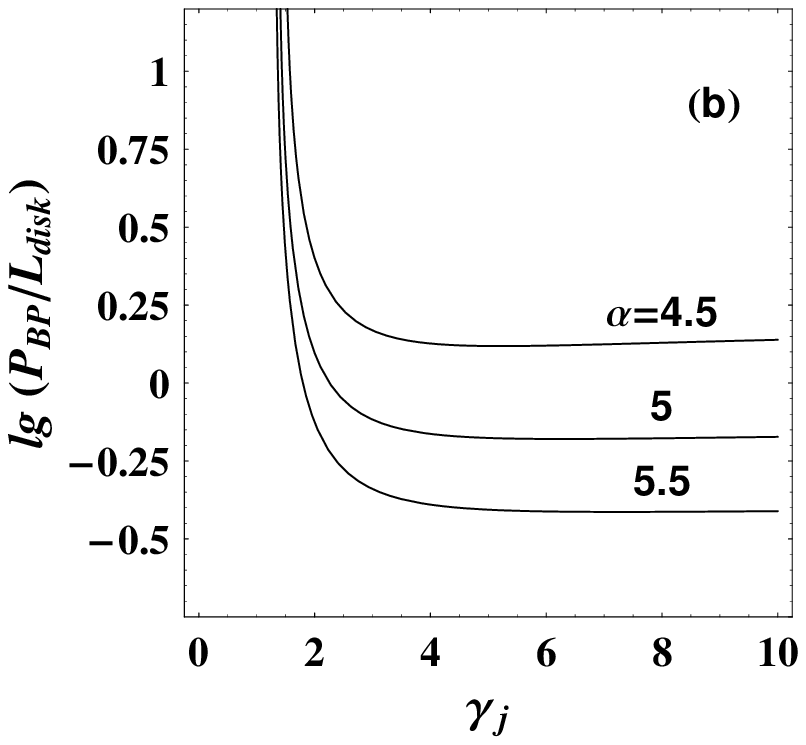}}
\caption[]{The curves of $\log \left( {{P_{BP} } \mathord{\left/
{\vphantom {{P_{BP} } {L_{disk} }}} \right.
\kern-\nulldelimiterspace} {L_{disk} }} \right)$ versus $\gamma_{j}
$ for different values of $\alpha $ with (a) $a_{*}= 0$ and (b)
$a_{*} = 0.95$.} \label{fig6}
\end{figure}

As shown in Fig. 6, the ratio ${P_{BP} } \mathord{\left/ {\vphantom
{{P_{BP} } {L_{disk} }}} \right. \kern-\nulldelimiterspace}
{L_{disk} }$ decreases with the Lorentz factors $\gamma _j $ very
steeply for $\gamma _j < 3$, while it almost remains constant for
$\gamma _j \ge 3$. And the ratio decreases with the increasing
self-similar index $\alpha $, while it increases with the black hole
spin $a_ * $. These results are consistent with those given in C02,
except that the ratios obtained in our model are greater than those
given in C02 for the same values of $a_ * $, $\alpha $, and
$\gamma_{j} $. The difference between the two models might arise
from the influence of the BP process on the disk radiation and the
contribution of Poynting flux in the jet being not taken into
account in C02.

McKinney (2006) proposes that AGNs are observed to have jets with
Lorentz factor $\sim $10 (Urry {\&} Padovani 1995; Biretta, Sparks
{\&} Macchetto 1999). Considering that ${P_{BP} } \mathord{\left/
{\vphantom {{P_{BP} } {L_{disk} }}} \right.
\kern-\nulldelimiterspace} {L_{disk} }$ is insensitive to the
Lorentz factor for $\gamma _j \ge 3$, we take $\gamma _j = 10$ when
studying the variation in ${P_{BP} } \mathord{\left/ {\vphantom
{{P_{BP} } {L_{disk} }}} \right. \kern-\nulldelimiterspace}
{L_{disk} }$ with the two parameters $a_ * $ and $\alpha $. As shown
in Fig. 7, the contours of ${P_{BP} } \mathord{\left/ {\vphantom
{{P_{BP} } {L_{disk} }}} \right. \kern-\nulldelimiterspace}
{L_{disk} }$= 0.1, 1, and 10, and the contour of $\left( {F_{rad} }
\right)_{\min } = \mbox{0}$ are plotted in $a_* - \alpha $ parameter
space, by which the parameter space are divided into regions I, II,
III, and IV.

In region I the disk luminosity dominates the BP power
significantly, while the BP power is comparable to the disk
luminosity in regions II and III, in which the radiation cooling
mode coexists with outflow/jet cooling mode. The BP power is less
and greater than the disk luminosity, in regions II and III,
respectively. Region IV is indicated as a forbidden region, which
corresponds to the negative radiation flux. The contour of ${P_{BP}
} \mathord{\left/ {\vphantom {{P_{BP} } {L_{disk} }}} \right.
\kern-\nulldelimiterspace} {L_{disk} } = 10$ lies in the region IV,
which means that the positive radiation flux requires the BP power
not to be much greater than the disk luminosity. Comparing with the
results given in C02, the larger self-similar index $\alpha $ is
required by the positive disk radiation flux.

\begin{figure}
\centering
\centerline{\includegraphics[width=70mm,height=70mm]{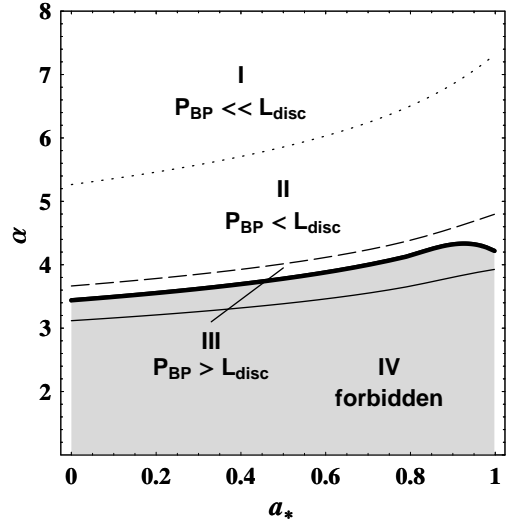}}
\caption[]{The $a_ * - \alpha $ parameter space: the contours of
${P_{BP} } \mathord{\left/ {\vphantom {{P_{BP} } {L_{disk} }}}
\right. \kern-\nulldelimiterspace} {L_{disk} }$ =0.1, 1, and 10
correspond respectively to the dotted, dashed and thin-solid lines,
respectively, and the contour of $\left( {F_{rad} } \right)_{\min }
= \mbox{0}$ is a thick solid line for $\gamma _j = 10$.}
\label{fig7}
\end{figure}

To fit the relativistic jet power from AGNs based on the magnetic
field configuration depicted in Fig.1, we should compare the
importance of the BZ power with respect to the BP power.
Incorporating Eqs. (\ref{eq5}) and (\ref{eq25}), we have the
contours of ${P_{BZ} } \mathord{\left/ {\vphantom {{P_{BZ} } {P_{BP}
}}} \right. \kern-\nulldelimiterspace} {P_{BP} } = $constant in the
$a_ * - \alpha $ parameter space with $\gamma _j = 10$ as shown in
Fig. 8, in which ${P_{BZ} } \mathord{\left/ {\vphantom {{P_{BZ} }
{P_{BP} }}} \right. \kern-\nulldelimiterspace} {P_{BP} } = $0.1, 1,
and 10, respectively. The shaded region below the thick solid line
represents the forbidden region as argued above. The region between
the dotted and thin solid lines indicates $0.1 < {P_{BZ} }
\mathord{\left/ {\vphantom {{P_{BZ} } {P_{BP} }}} \right.
\kern-\nulldelimiterspace} {P_{BP} } < 10$, i.e., the BZ power is
comparable to the BP power. The region above the thin solid line and
the one below the dotted line represent $P_{BP} < < P_{BZ} $ and
$P_{BP}
> > P_{BZ} $, respectively. Thus the contribution of the BZ and BP
processes to the jet power from AGNs can be determined by the values
of the parameters $\alpha $ and $a_ * $.

\begin{figure}
\centering
\centerline{\includegraphics[width=70mm,height=70mm]{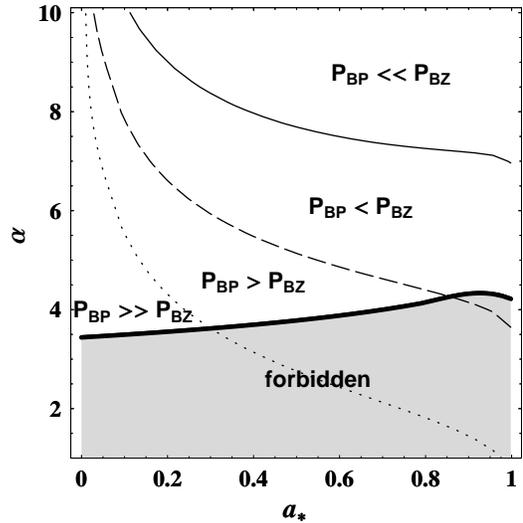}}
\caption[]{The $a_ * - \alpha $ parameter space: the contours of
${P_{BZ} } \mathord{\left/ {\vphantom {{P_{BZ} } {P_{BP} }}} \right.
\kern-\nulldelimiterspace} {P_{BP} } = $0.1, 1, and 10 correspond
respectively to the dotted, dashed, and thin-solid lines, and the
contour of $\left( {F_{rad} } \right)_{\min } = \mbox{0}$ is a thick
solid line for $\gamma _j = 10$.} \label{fig8}
\end{figure}

Based on the above discussion, we find that both the BZ and BP
powers should be taken into account in fitting the relativistic jet
power, provided that the values of $a_ * $ and $\alpha $ are taken
in the region between the dotted and thin-solid lines as shown in
Fig. 8. Thus the jet power can be fitted as the sum of the BZ and BP
powers,

\begin{equation}
\label{eq26} Q_{jet} = P_{BP} + P_{BZ} .
\end{equation}

\noindent To study the relationship between the jet power and the
disk luminosity, we plot the contours of ${Q_{jet} } \mathord{\left/
{\vphantom {{Q_{jet} } {L_{disk} }}} \right.
\kern-\nulldelimiterspace} {L_{disk} }$=constant in $a_ * - \alpha $
parameter space with $\gamma _j = 10$ as shown in Fig. 9.

\begin{figure}
\centering
\centerline{\includegraphics[width=70mm,height=70mm]{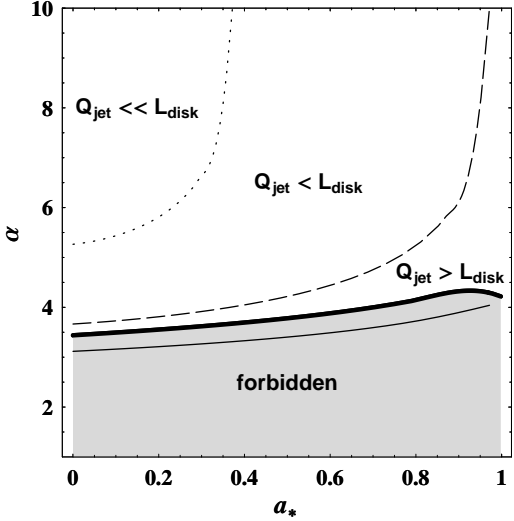}}
\caption[]{The $a_ * - \alpha $ parameter space- The contours of
${Q_{jet} } \mathord{\left/ {\vphantom {{Q_{jet} } {L_{disk} }}}
\right. \kern-\nulldelimiterspace} {L_{disk} } = $0.1, 1 and 10
correspond respectively to the dotted, dashed, and thin-solid lines,
and the contour of $\left( {F_{rad} } \right)_{\min } = \mbox{0}$ is
a thick solid line for $\gamma _j = 10$.} \label{fig9}
\end{figure}

It is found from Fig. 9 that the jet power and the disk luminosity
are comparable, except that the black hole rotates slowly and the
magnetic field decreases very steeply with the cylindrical radius.
The region above the thin solid line in the $a_ * - \alpha $
parameter space represents $Q_{jet} < < L_{disk} $. As shown in Fig.
9, the contour of ${Q_{jet} } \mathord{\left/ {\vphantom {{Q_{jet} }
{L_{disk} }}} \right. \kern-\nulldelimiterspace} {L_{disk} } = 10$
lies in the forbidden region, and it implies that the jet power
cannot be significantly greater than the disk luminosity, which is
required by non-negative disk flux.

According to MT03 the disk luminosity is related to the broad-line
region luminosity by $L_{BLR} \approx 0.1L_{disk} $. Taking the
accretion rate at ISCO as $\dot {M}_{acc} (r_{ms} ) = 0.1\dot
{M}_{Edd} $\textbf{,} we derive the jet power and broad-line region
luminosity of 11 FSRQs and 17 SSRQs based on Eqs. (\ref{eq5}),
(\ref{eq24}), (\ref{eq25}), and (\ref{eq26}) as shown in Table 1. In
addition, we mark the points of values of $a_\ast $ and $\alpha $
corresponding to these sources in the $a_ * - \alpha $ parameter
space as shown in Fig. 10.

\begin{table*}
\caption{The values of the concerned parameters for fitting the jet
power and broad-line region luminosity and black hole mass of 11
FSRQs and 17 SSRQs, where $\gamma _j = 10$ and $\dot {M}_{acc}
(r_{ms} ) = 0.1\dot {M}_{Edd} $ are assumed. }
 \label{table:1}
\centering
\begin{tabular}{c c c c c c c c } \hline \hline
Sources & $\log Q_{jet} $  & $\log L_{BLR} $ & $\log M_{BH} $&
$a_\ast $ & $\alpha $& $\log P_{BZ} $ & $\log P_{BP} $  \\
\hline 0017+154$^{ S}$& 47.19 $^{[BM87]}$& 45.79 $^{[CIV,C91]}$&
9.62 $^{[MgII,H02]}$& 0.89& 4.51& 46.94&
46.83 \\
\hline 0022+297$^{ S}$& 45.61 $^{[K98]}$& 44.15 $^{[H\beta
,S93]}$& 7.81 $^{[H\beta ,G01]}$& 0.98& 4.81& 45.44&
45.13 \\
\hline 0119-046$^{ S}$& 46.52 $^{[R99]}$& 46.00 $^{[MgII,SS91]}$&
9.91 $^{[MgII,B94]}$& 0.25& 4.64& 45.74&
46.44 \\
\hline 0134+329$^{ S}$& 46.30 $^{[BM87]}$& 44.96$^{[H\beta
,JB91]}$& 8.65 $^{[H\beta ,C97]}$& 0.94& 4.90& 46.12&
45.83 \\
\hline 0238+100$^{ S}$& 46.10 $^{[N89]}$& 45.59 $^{[CIV,C91]}$&
9.48 $^{[CIV,C91]}$& 0.29& 4.76& 45.46&
45.99 \\
\hline 0336-019$^{ F}$& 45.18 $^{[BM87]}$& 45.00 $^{[H\beta
,JB91]}$& 8.89 $^{[H\beta ,G01]}$& 0.13& 5.12& 44.18&
45.13 \\
\hline 0403-132$^{ F}$& 45.60 $^{[BM87]}$& 45.25 $^{[H\beta
,O84]}$& 9.08 $^{[H\beta ,M96]}$& 0.35& 5.40& 45.23&
45.36 \\
\hline 0607-157$^{ F}$& 44.20 $^{[BM87]}$& 43.56 $^{[H\beta
,H78]}$& 7.32 $^{[H\beta ,G01]}$& 0.58& 5.55& 44.01&
43.75 \\
\hline 0637-752$^{ F}$& 46.48 $^{[CJ01]}$& 45.44 $^{[H\beta
,T93]}$& 8.81 $^{[H\beta ,G01]}$& 0.98& 9.15& 46.47&
44.85 \\
\hline 0837-120$^{ S}$& 44.79 $^{[R99]}$& 45.00 $^{[H\beta
,B96]}$& 8.86 $^{[H\beta ,B96]}$& 0.12& 6.01& 44.06&
44.70 \\
\hline 0838+133$^{ F}$& 46.19 $^{[BM87]}$& 45.14 $^{[H\beta
,JB91]}$& 8.67 $^{[H\beta ,B96]}$& 0.94& 6.38& 46.12&
45.36 \\
\hline 0903+169$^{ S}$& 45.30 $^{[BM87]}$& 44.69 $^{[H\beta
,B96]}$& 8.39 $^{[H\beta ,B96]}$& 0.65& 6.46& 45.20&
44.56 \\
\hline 1023+067$^{ S}$& 46.50 $^{[BM87]}$& 45.07 $^{[MgII,C91]}$&
8.99 $^{[MgII,C91]}$& 0.85& 4.29& 46.20&
46.20 \\
\hline 1040+123$^{ S}$& 46.27 $^{[BM87]}$& 45.11 $^{[MgII,N79]}$&
8.76 $^{[MgII,H02]}$& 0.91& 5.29& 46.12&
45.73 \\
\hline 1250+568$^{ S}$& 45.50 $^{[BM87]}$& 44.57 $^{[H\beta
,JB91]}$& 8.31 $^{[H\beta ,B96]}$& 0.74& 5.18& 45.30&
45.06 \\
\hline 1253-055$^{ F}$& 45.70 $^{[BM87]}$& 44.64 $^{[H\beta
,M96]}$& 8.28 $^{[H\beta ,G01]}$& 0.88& 5.54& 45.11&
45.57 \\
\hline 1318+113$^{ S}$& 46.86 $^{[G91]}$& 45.86 $^{[CIV,C91]}$&
9.32 $^{[CIV,C91]}$& 0.95& 7.80& 46.83&
45.63 \\
\hline 1334-127$^{ F}$& 44.91 $^{[CJ01]}$& 44.18 $^{[MgII,S93]}$&
7.98 $^{[MgII,W86]}$& 0.57& 5.08& 44.64&
44.58 \\
\hline 1442+101$^{ S}$& 46.95 $^{[BM87]}$& 45.93 $^{[CIV,C91]}$&
9.93 $^{[H\beta ,H03]}$& 0.44& 4.08& 46.31&
46.84 \\
\hline 1559+173$^{ S}$& 46.81 $^{[S90]}$& 45.66 $^{[MgII,C91]}$&
9.25 $^{[MgII,C91]}$& 0.93& 5.59& 46.69&
46.18 \\
\hline 1606+289$^{ S}$& 46.56 $^{[BM87]}$& 45.61 $^{[CIV,C91]}$&
9.37 $^{[CIV,C91]}$& 0.73& 5.04& 46.34&
46.16 \\
\hline 1622+238$^{ S}$& 46.48 $^{[BM87]}$& 45.34 $^{[MgII,SS91]}$&
9.53 $^{[H\beta ,B96]}$& 0.16& 3.63& 44.97&
46.47 \\
\hline 1641+399$^{ F}$& 45.30$^{ [BM87]}$& 45.47 $^{[H\beta
,L96]}$& 9.27 $^{[H\beta ,M96]}$& 0.29& 8.07& 45.24&
44.39 \\
\hline 1954-388$^{ F}$& 44.12 $^{[CJ01]}$& 44.20 $^{[H\beta
,T93]}$& 7.99 $^{[H\beta ,O02]}$& 0.32& 8.11& 44.07&
43.14 \\
\hline 1954+513 $^{F}$& 46.04 $^{[K90]}$& 45.39 $^{[Mg?,L96]}$&
9.18 $^{[MgII,L96]}$& 0.54& 5.27& 45.79&
45.68 \\
\hline 1655+077$^{ F}$& 45.00 $^{[M93]}$& 43.62 $^{[MgII,W86]}$&
7.28 $^{[MgII,W86]}$& 0.96& 4.92& 44.83&
44.51 \\
\hline 2120+168$^{ S}$& 46.88 $^{[BM87]}$& 45.57 $^{[CIV,O94]}$&
9.68 $^{[MgII,H02]}$& 0.54& 3.85& 46.28&
46.76 \\
\hline 2354+144$^{ S}$& 46.73 $^{[H83]}$& 44.75 $^{[H\beta
,C91]}$& 9.37 $^{[H\beta ,C91]}$& 0.87& 5.91& 46.63&
46.06 \\
\hline
\end{tabular}
\label{tab1}

\textbf{Notes:   }Column   (\ref{eq1}):   IAU   source  name.  The
superscript   ``F''   and   ``S''   represent   FSRQs  and  SSRQs,
respectively.  Column (\ref{eq2}): jet power Q$_{jet}$ in units of
erg  s$^{  -  1}$.  The superscript represents references of radio
extended  flux  density  in  calculating  the  jet  power.  Column
(\ref{eq3}):  broad-line region luminosity in units of erg s$^{ -
1}$.  The  superscript represents the adopted lines in calculating
broad-line  region  luminosity  and  references for lines. Column
(\ref{eq4}):  black  hole  mass  in  units  of  M$_{ \odot }$. The
superscript  represents  lines  for estimating black hole mass and
references  for  lines.  Column (\ref{eq5}): the fitted black hole
spin.  Column  (\ref{eq6}):  the fitted self-similar index. Column
(\ref{eq7}):  the  corresponding  BZ  power in units of erg s$^{ -
1}$.  Column (\ref{eq8}): the corresponding BP power in units of e
erg                  s$^{                  -                  1}$.

\textbf{References:}   B94:   Brotherton   et   al.  (1994).  B96:
Brotherton  (1996).  BM87:  Browne {\&} Murphy (1987). C91: Corbin
(1991).  C97:  Corbin (1997). CJ01: Cao {\&} Jiang (2001). G01: Gu
et  al.  (2001).  H78: Hunstead et al. (1978). H83: Hintzen et al.
(1983).  H02: Hough et al. (2002). H03: Hirst et al. (2003). JB91:
Jackson  {\&}  Browne  (1991).  K90: Kollgaard et al. (1990). K98:
Kapahi  et al. (1998). L96: Lawrence et al. (1996). M93: Murphy et
al.  (1993).  M96:  Marziani et al. (1996). N79: Neugebauer et al.
(1979).  N89:  Neff  et  al.  (1989). O84: Oke et al. (1984). O94:
Osmer  et al. (1994). O02: Oshlack et al. (2002). R99: Reid et al.
(1999).  S90:  Saikia  et  al. (1990). S93: Stickel et al. (1993).
SS91:  Steidel  {\&} Sargent (1991). T93: Tadhunter et al. (1993).
W86:                         Wilkes                        (1986).
\end{table*}

\begin{figure}
\centering
\centerline{\includegraphics[width=70mm,height=70mm]{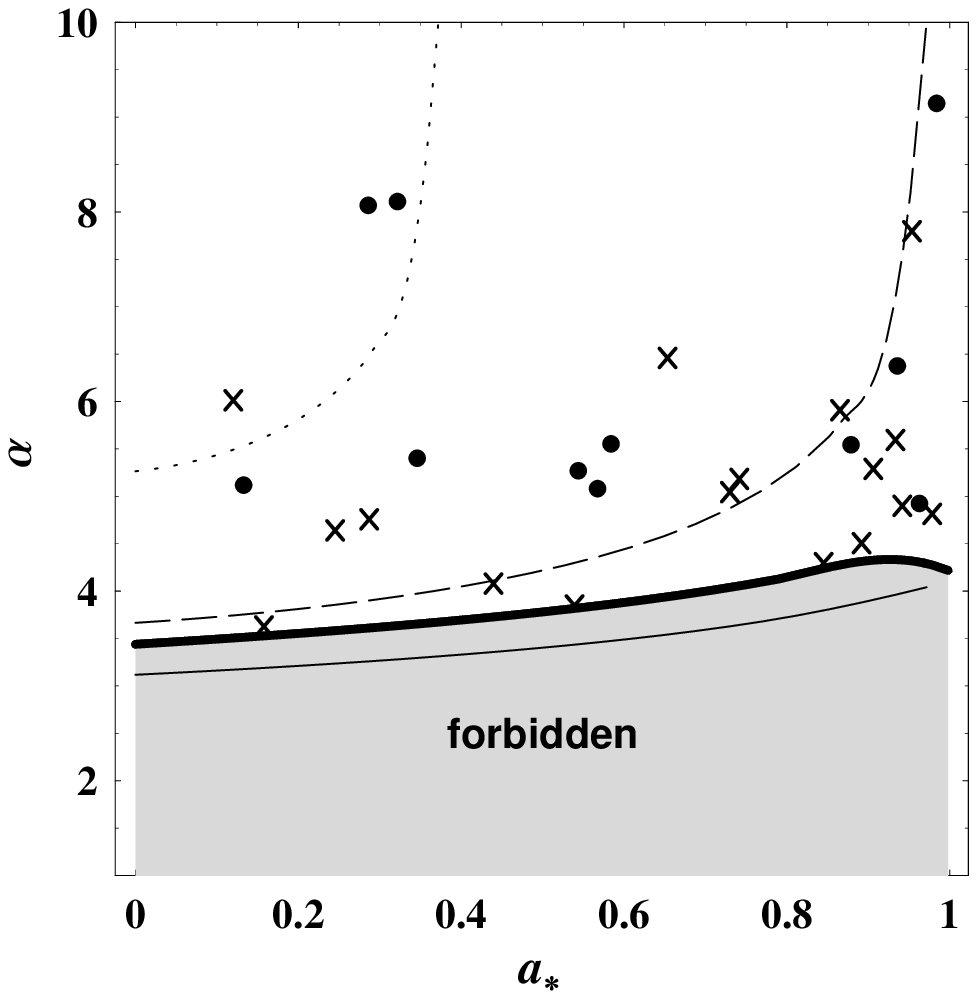}}
\caption[]{The $a_ * - \alpha $ parameter space- The values of the
parameters $a_\ast $ and $\alpha $ correspond to the jet power and
the disk luminosity of 11 FSRQs (solid circles), 17 SSRQs (crosses)
listed in Table 1, and the contours of ${Q_{jet} } \mathord{\left/
{\vphantom {{Q_{jet} } {L_{disk} }}} \right.
\kern-\nulldelimiterspace} {L_{disk} } = $0.1, 1, and 10 correspond
respectively to the dotted, dot-dashed, dashed, and thin-solid
lines, and the contour of $\left( {F_{rad} } \right)_{\min } =
\mbox{0}$ is a thick solid line for $\gamma _j = 10$.} \label{fig10}
\end{figure}

It is found from Fig. 10 that the parameters $a_\ast $ and $\alpha $
of 25 samples (except one SSRQ and two FSRQs) fall in the region
between the dotted and thick-solid lines, which indicates that the
jet power is comparable to the disk luminosity. This result is
consistent with what was derived by MT03 and D'Elia, Padovani {\&}
Landt (2003). Other sources given in L06 cannot be fitted by our
model for the following reasons: (\ref{eq1}) the constraint of
non-negative disk radiation flux expressed by Eq. (\ref{eq23}), and
(\ref{eq2}) the accretion rate at ISCO is assumed to be 0.1
Eddington accretion rate. Some sources with disk luminosity
exceeding this limit are not fitted clearly.

\section{ CONCLUSIONS}

The coupling of disk-jet is essential for jet production from AGNs
and stellar black hole systems (Blandford 2002). The BZ and BP
processes are two main mechanisms for driving the jets. In this
paper a simplified model of jet power from AGNs is proposed by
combining the BZ and BP processes with the disk accretion. The
expressions of the BP power and disk luminosity can be derived from
the conservation laws of mass, angular momentum, and energy. These
equations consist of a closed set for resolving the BP power and the
disk luminosity. It turns out that the relative importance of a
series of quantities related to the jet power, such as the BP power,
the BZ power, the disk luminosity, and the jet power itself can be
displayed visually in the $a_\ast - \alpha $ parameter space.

It is shown that the disk radiation flux and luminosity decrease due
to a fraction of the accretion energy being channelled into the
outflow/jet in the BP process, and the dominant cooling mode of the
accretion disk is determined mainly by how the the poloidal magnetic
field decreases with the cylindrical radius of the jet. The dominant
mode is radiation cooling for the magnetic field decreasing very
steeply with the cylindrical radius, while the mode could be
outflow/jet cooling for the magnetic fields decreasing less steeply.
However, the strength of the outflow/jet cooling should be less than
some critical values to avoid a negative disk radiation.

In this model the jet power is regarded as the sum of the BZ and BP
powers, which are related to the disk accretion rate by assuming a
relation connecting the magnetic field at the black hole horizon
with the accretion rate at ISCO [given by Eq. (\ref{eq4})], and the
broad-line region luminosity is assumed to be one tenth of the disk
luminosity. Based on these assumptions, we fit the jet power and
broad-line region luminosity of 11 FSRQs and 17 SSRQs, whose jet
power is almost the same as the disk luminosity.

In this simplified model, the poloidal magnetic field is assumed to
be anchored in a thin disk, scaling with the disk radius in a power
law, and the variation of the poloidal magnetic with the cylindrical
radius is described by a self-similar index given by C02. As a
matter of fact, the magnetic field configurations could be much more
complicated, and the accretion mode can affect the jet power and
disk luminosity significantly. We shall improve our model by
combining different magnetic field configurations with inefficient
accretion mode, such as ADAF, to fit the observations of the AGNs
with strongly dominated jet power in our future work.

\begin{acknowledgements}
This work is supported by the National Natural Science Foundation
of China under grants 10573006.
\end{acknowledgements}

{}

\end{document}